\documentclass[runningheads,a4paper]{llncs}

\usepackage{amssymb}
\usepackage{graphicx}
\usepackage[ruled,vlined,linesnumbered]{algorithm2e}
\usepackage{float}
\usepackage{color}
\usepackage{soul}
\usepackage{amsmath}
\usepackage{caption}
\usepackage{makecell}
\usepackage{enumitem} 

\usepackage{url}

\urldef{\mailsa}\path|{msarker@swin.edu.au}|
\urldef{\mailsb}\path|{flora.salim@rmit.edu.au}|
\urldef{\mailsc}\path|erika.siebert-cole, peter.strasser, lncs}@springer.com|    
\newcommand{\keywords}[1]{\par\addvspace\baselineskip
\noindent\keywordname\enspace\ignorespaces#1}

\begin{document}

\mainmatter  

\title{Mining User Behavioral Rules from Smartphone Data through Association Analysis}

\titlerunning{Behavioral association analysis}

%
%

\author{Iqbal H. Sarker$^1$%
\thanks{Springer International Publishing (preprint version), The 22nd Pacific-Asia Conference on Knowledge Discovery and Data Mining (PAKDD), June 3rd - 6th, 2018, Melbourne, Australia.}%
\and Flora D. Salim$^2$}
\authorrunning{Iqbal H. Sarker et al.}

\institute{\textsuperscript 1 Department of Computer Science \& Software Engineering,\\
	School of Software and Electrical Engineering,\\
	Swinburne University of Technology,
	Melbourne, Australia.\\
	\mailsa\\
	\textsuperscript 2 School of Science (CS \& IT), \\
	RMIT University, Melbourne, Australia.\\
	\mailsb\\
}

%
%

\toctitle{Lecture Notes in Computer Science}
\tocauthor{Authors' Instructions}
\maketitle

\begin{abstract}
The increasing popularity of smart mobile phones and their powerful sensing capabilities have enabled the collection of rich contextual information and mobile phone usage records through the device logs. This paper formulates the problem of mining \textit{behavioral association rules} of individual mobile phone users utilizing their smartphone data. Association rule learning is the most popular technique to discover rules utilizing large datasets. However, it is well-known that a large proportion of association rules generated are \textit{redundant}. This redundant production makes not only the rule-set unnecessarily large but also makes the decision making process more complex and ineffective. In this paper, we propose an approach that effectively identifies the \textit{redundancy} in associations and extracts a concise set of \textit{behavioral association rules} that are non-redundant. The effectiveness of the proposed approach is examined by considering the real mobile phone datasets of individual users.

\end{abstract}

\section{Introduction}
Now-a-days, mobile phones have become part of our life. The number of mobile cellular subscriptions is almost equal to the number of people on the planet \cite{pejovic2014interruptme}. The phones are, for most of the day, with their owners as they go through their daily routines. People use smart mobile phones for various activities such as voice communication, Internet browsing, apps using, e-mail, online social network, instant messaging etc. \cite{pejovic2014interruptme}. 

The sensing capabilities of smart mobile phones have enabled the collection of rich contextual information and mobile phone usage records through the device logs \cite{zhu2014mining}. These are phone call logs \cite{sarker2016phone}, app usages logs \cite{srinivasan2014mobileminer}, mobile notification logs \cite{mehrotra2016prefminer}, web logs \cite{halvey2005time}, context logs \cite{zhu2014mining} etc. The discovered behavioral association rules from such mobile phone data, can be used for building the adaptive, intelligent and context-aware personalized systems, such as smart interruption management system, intelligent mobile recommender system, context-aware smart searching, and various predictive services, in order to assist them intelligently in their daily activities in a context-aware pervasive computing environment.

In this paper, we mainly focus on \textit{mining individual's phone call behavior} $(Accept | Reject | Missed | Outgoing)$ utilizing their phone log data. In the real-world, mobile phone users' behaviors are not identical to all. Individual user may behave differently in different contexts. Let's consider a smart phone call handling service, a mobile phone user typically `rejects' the incoming phone calls, if s/he is in a `meeting'; however, `accepts' if the call comes from his/her `boss'. Hence, [reject, accept] are the user phone call behaviors, and [meeting, boss] are the associated contexts that have a strong influence on users to make decisions. Context is defined as \textit{``any information that can be used to characterize the situation of a user"}, such as temporal (e.g., day, time), social activity or situation (e.g., meeting), location (e.g., office), social relationship (e.g., boss) etc. In this work, we aim to extract a concise set of \textit{behavioral association rules} that are \textit{non-redundant}, expressing an individual's phone call behavior in such multi-dimensional contexts for a particular confidence threshold preferred by individuals. The setting of this threshold for creating rules will vary according to an \textit{individual's preference} as to how interventionist they want the call handling agent to be. Let's consider an example, one person  may want the agent to reject calls where in the past he/she has rejected calls more than, say, 95\% of the time - that is, at a threshold of 95\%. Another individual, on the other hand, may only want the agent to intervene if he/she has rejected calls in, say, 80\% of past instances. Such preferences may vary from user-to-user in the real world. 
 
In the area of mobile data mining, association rule learning \cite{agrawal1994fast} is the most common techniques to discover rules of mobile phone users. In particular, a number of researchers \cite{mehrotra2016prefminer,srinivasan2014mobileminer,zhu2014mining} have used association rule learning to mine rules capturing mobile phone users' behavior for various purposes. However, the drawback is - Association rule learning technique discovers all associations of contexts in the dataset that satisfy the user specified minimum support and minimum confidence constraints. As a result, it produces a huge number of \textit{redundant rules} (affects the quality and usefulness of the rules) because of considering all possible combinations of contexts without any intelligence. According to \cite{fournier2012mining}, association rule learning technique produces up to 83\% redundant rules that makes the rule-set unnecessarily large. Therefore, it is very difficult for the decision making agents to determine the most interesting ones and consequently makes the decision making process ineffective and more complex.

In this paper, we address the above mentioned issues and propose an approach that effectively identifies the \textit{redundancy} in associations and extracts a concise set of \textit{behavioral association rules} that are \textit{non-redundant}. In our approach, we first design an association generation tree, in which each branch denotes a test on a specific context value determining according to the precedence of contexts, and each corresponding node either interior or leaf represents the outcome, including the identified `REDUNDANT' nodes, for the test. Once the tree has been generated, we extract rules by traversing the tree from root node to each rule producing node that satisfies the user preferred confidence threshold. 

The contributions are summarized as follows:

\begin{itemize}  
	\item We effectively identify the \textit{redundancy in associations} while producing rules rather than in post-processing.
		
	\item We propose an approach that extracts a concise set of \textit{behavioral association rules} that are \textit{non-redundant}.
	
	\item We have conducted experiments on real mobile phone datasets to show the effectiveness of our approach comparing with traditional association rule learning algorithm.
\end{itemize}

The rest of the paper is organized as follows. Section \ref{Background} reviews the background of association rule learning techniques. We discuss the redundancy in associations in Section \ref{redundancy}. Section \ref{Our Approach} presents our approach. We report the experimental results in Section \ref{Experiments}. Finally, Section \ref{Conclusion and Future Work} concludes this paper highlighting the future work.

 \section{Association Rules: A Background}
 \label{Background}
 Association rule mining is one of the most important and well researched techniques in data mining. In this section, we introduce some basic and classic approaches for association rule mining. An association rule is an implication in the form of $A \Rightarrow C$, where, $A$ is called antecedent while $C$ is called consequent, the rule means $A$ implies $C$.
 
 The AIS algorithm, proposed by Agrawal et al. \cite{agrawal1993mining}, is the first algorithm designed for association rule mining. The main drawback of the AIS algorithm is too many candidate itemsets that finally turned out to be small are generated, which requires more space and wastes much effort that turned out to be useless. At the same time this algorithm requires too many passes over the whole database. The SETM algorithm proposed by \cite{houtsma1995set} exhibits good performance and stable behavior, with execution time almost insensitive to the chosen minimum support but has the same disadvantage of the AIS algorithm.
 
 Apriori, Aprioiri-TID and Apriori-Hybrid algorithms are proposed by Agrawal in \cite{agrawal1994fast}. The performance is these algorithms are better than AIS and SETM. The Apriori algorithm takes advantage of the fact that any subset of a frequent itemset is also a frequent itemset. The algorithm can therefore, reduce the number of candidates being considered by only exploring the itemsets whose support count is greater than the minimum support count. All infrequent itemsets can be pruned if it has an infrequent subset. Apriori-TID and Apriori-Hybrid are designed based on Apriori algorithm. Another algorithm Predictive Apriori proposed by Scheffer \cite{scheffer2005finding} generates rules by predicting accuracy combining from support and confidence. So sometimes it produced the rules with large support but low confidence and got unexpected results.
  
 Han et al. \cite{han2000mining} have designed a tree based rule mining algorithm FP-Tree. However, FP-Tree is difficult to be used in an interactive mining system. During the interactive mining process, users may change the threshold of support according to the rules. The changing of support may lead to repetition of the whole mining process. Das et al. \cite{das2001rapid} have designed another tree based association rule mining method RARM that uses the tree structure to represent the original database and avoids candidate generation process. RARM is claimed to be much faster than FP-Tree algorithm but also faces the same problem of FP-tree \cite{zhao2003association}. Flach et. al \cite{flach2001confirmation} introduces an approach with learning first-order logic rules. This algorithm is able to deal with explicit negation. However, this algorithm can not learn rules in case of depth search.
  
 Among the association rule mining algorithms, Apriori \cite{agrawal1994fast} is a great improvement in the history of association rule mining \cite{zhao2003association}. This is the most popular and common algorithm for mining association rules. The key strength of association rule mining is it's completeness. It finds all associations in the data that satisfy the user specified constraints. However, the main drawback is that - it produces a huge number of \textit{redundant} associations, that makes the behavior modeling approach ineffective for mobile phone users.
 
 Unlike these works, in this paper, we propose an approach that effectively identifies the \textit{redundancy} in associations and extracts a concise set of \textit{behavioral association rules} that are \textit{non-redundant} for individual mobile phone users utilizing their mobile phone data. 
 
 \section{Redundancy in Association Rules}
 \label{redundancy}
 Association rule learning algorithms produce many rules $(A \Rightarrow C)$ that have common consequent $(C)$ `behavior' but different antecedent $(A)$ `contexts'. Indeed many of those antecedent contexts are proper subset of others rules.
 
 Let, two rules $R_1: A_1 \Rightarrow C_1$ and $R_2: A_2 \Rightarrow C_2$, we call the latter one redundant with the former one if $A_1 \subseteq A_2$ and $C_1 = C_2$. From this definition of redundancy, if we have a general rule $R_g: A_1 \Rightarrow C_1$ and there is no other more specific rule $A_1B_1 \Rightarrow C_2$ in existence such that confidence of $A_1B_1 \Rightarrow C_2$ is equal or larger than the confidence of $R_g: A_1 \Rightarrow C_1$ and $A_1 \subseteq A_1B_1$, $C_1 = C_2$, then the rule $A_1B_1 \Rightarrow C_2$ is said to be non-redundant with $R_g: A_1 \Rightarrow C_1$.
 	
 For example, typically a user rejects most of the incoming calls (83\%), when she is in a meeting, i.g., the rule is $(meeting \Rightarrow reject)$ [say, user preferred confidence threshold 80\%]. Another example is, the user rejects most of the incoming calls (90\%) of her friends, when she is in a meeting, i.g., the rule is $(meeting, friend \Rightarrow reject)$. Both rules are valid in terms of confidence as the rules satisfy the user preferred confidence threshold. However, the later one is considered as redundant rule as the former one is able to take the same decision with minimal number of contexts. Additional context can play a significant role if it reflects different behavior. Table \ref{sample-rules} shows an example of a set of association rules and their non-redundant production for a preferred minimum confidence 80\%. According to Table \ref{sample-rules} $R_2, R_3, R_4, R_5$ are redundant rules as only $R_1$ is able to take the same decision with minimal number of contexts. On the other hand, $R_1$ and $R_6$ are considered as non-redundant rule, in which we are interested in.
 
		\begin{table*}
	 	 	\centering
	 	 	\caption{Sample traditional association rules and corresponding non-redundant behavioral association rules of a sample user.}
	 	 	\label{sample-rules}
	 	 	\begin{tabular}{|c|c|} 
	 	 		\hline
	 	 		\bf \makecell{Association Rules \\ (Traditional)} & \bf \makecell{Association Rules \\ (Non-redundant)} \\  
	 	 		\hline
	 	 		$\begin{aligned} 
	 	 			
	 	 			R_1:  Meeting \Rightarrow Reject \\ (conf = 83\%) \\
	 	 			R_2:  Meeting, Friend \Rightarrow Reject \\ (conf = 90\%) \\
	 	 			R_3:  Meeting, Colleague \Rightarrow Reject \\ (conf = 88\%) \\
	 	 			R_4:  Meeting, Friend, Monday[t1] \Rightarrow Reject \\ (conf = 100\%) \\
	 	 			R_5:  Meeting, Colleague, Friday[t2] \Rightarrow Reject \\ (conf = 98\%) \\
	 	 			R_6:  Meeting, Boss \Rightarrow Accept \\ (conf = 100\%) \\
			 	 		 	 		
	 	 		\end{aligned}$
	 	 		& 
	 	 		$\begin{aligned} 
	 	 			R_1:  Meeting \Rightarrow Reject \\ (conf = 83\%) \\ 
	 	 			R_6:  Meeting, Boss \Rightarrow Accept \\ (conf = 100\%) \\	 
				\end{aligned}$ \\ 
			 	 	 			 
	 	 	\hline 
	 	 	\end{tabular}
\end{table*}
 
 \section{Our Approach}
 \label{Our Approach}
 In this section, we present our approach for mining behavioral association rules of individual mobile user behavior utilizing their mobile phone data.
 
 \subsection{Association Generation Tree (AGT)}
 In this first step, we generate a tree based on multi-dimensional contexts and corresponding usage behavior of mobile phone users. As different contexts might have differing impacts in behavioral rules, we identify the precedence of contexts in a dataset while generating the tree.
 
 \subsubsection{Identifying the Precedence of Contexts:}
 In order to identify the precedence of contexts in a dataset, we calculate information gain which is a statistical property that measures how well a given attribute separates training examples into targeted behavior classes. The one with the highest information is considered as the highest precedence context. In order to define information gain precisely, we need to define entropy first. 
 
 Entropy is a measure of disorder or impurity. The entropy characterizes the impurity of an arbitrary collection of examples. It reaches it's maximum when the uncertainty is at a maximum and vice-versa. Formally entropy is defined as \cite{quinlan1993c4}:
 
 $$H(S)= -\sum_{x \in X} p(x)log_2p(x)$$
 
 Where, S is the current data set for which entropy is being calculated, X represents a set of classes in S, p(x) is the proportion of the number of elements in class x to the number of elements in set S.
 
 Information gain (IG) measures how much ``information" a feature gives us about the class. It is the expected reduction in entropy caused by partitioning the examples according to a given attribute. Features that perfectly partition should give maximal information. Unrelated features should give no information. To decide which attribute should be tested first, we find the one with the highest information gain. The formal definition of information gain is \cite{quinlan1993c4}-
 
 $$IG(A,S)= H(S)-\sum_{t \in T} p(t)H(t)$$
 
 Where, H(S) is the entropy of set S, T represents the subsets created from splitting set S by attribute A such that $S=\cup_{t \in T} t$, p(t) is the proportion of the number of elements in t to the number of elements in set S, H(t) is the entropy of subset t. 
 
 Let's consider a sample dataset of three different contexts and corresponding call response behavior of a mobile phone user X. For example, the contexts might be ranked as follows: \\
 
 \noindent $Rank1: Social \; Activity / Situation (S)\in \{meeting, lecture, lunch\}$ \\ 
 $Rank2: Social \; Relationship (R) \in \{boss, colleague, friend, unknown\}$ \\ 
 $Rank3: Temporal (T)\in \{time$-$of$-$the$-$week\}$ \\
 Where, \\
 $User \; phone  \; call  \; behavior (BH)\in \{Accept, Reject, Missed\}$ \\
 
 \subsubsection{Tree Generating Procedure and Extracting Non-redundant Rules:}
 A tree is a structure that includes a root node, branches, interior and/or leaf nodes \cite{sarker2017anapproach}. Each branch denotes a test on a specific context value, and each node (interior or leaf) denotes the outcome containing the behavior class with confidence value of the test. 
 
 To build tree, we follow a top-down approach, starting from a root node. The tree is partitioned into classes distinguished by the values of the most relevant context according to the precedence. Once the root node of the tree has been determined, the child nodes and it's arcs are created and added to the tree with the associated contexts and corresponding behavior with confidence value. While creating a node, we check whether it is redundant (`REDUNDANT' node) or not. 
 
 \textit{``A child node in the tree is called `REDUNDANT' node, if both the child node and it's parent node contain same behavior class and satisfy individual's preferred confidence threshold''}. 
 
 The algorithm recursively add new subtrees to each branching arc by adding child node one by one. If a node has 100\% (maximum) confidence then there is no need to elaborate it's children, otherwise we continue this process according to the number of contexts in the datasets. The final result is a multi-level tree with various nodes including `REDUNDANT' node according to their associated contexts. The overall process for constructing the tree is set out in Algorithm \ref{alg:tree}.
 
 \begin{algorithm}
 	 	\caption{Association Generation Tree}
 	 	\label{alg:tree}
 	 	\SetKwInOut{Data}{Data}
 	 	\SetAlgoLined
 	 	\Data{Dataset: $DS = {X_1,X_2,...,X_n}$ // each instance $X_i$ contains a number of nominal context-values and corresponding behavior class $BH$, confidence threshold = $t$}
 	 	
 	 	\KwResult{An association generation tree}
 	 	
 	 	\BlankLine
 	 	
 	 	\underline{Procedure AGT} $(DS, context \textunderscore list, BHs)$\;
 	 	
 	 	$N \leftarrow createNode()$ //create a root node for the tree \\
 	 	
 	 	\If{all instances in $DS$ belong to the same behavior class $BH$}
 	 	{
 	 		return $N$ as a leaf node labeled $BH$ with 100\% confidence. \\
 	 	}
 	 	
 	 	\If{context \textunderscore list is empty}
 	 	{
 	 		return $N$ as a leaf node labeled with the dominant behavior class and corresponding confidence value. \\
 	 	}
 		
		identify the highest precedence context $C_{split}$ for splitting and assign $C_{split}$ to the node $N$. \\
 	 
 	 	\ForEach{context value $val \in C_{split}$}
 	 	{  
 	 		create subset $DS_{sub}$ of $DS$ containing $val$. \\
 	 		
 	 		\If{$DS_{sub} \ne \phi$}
 	 		{
				identify the dominant behavior and calculate the confidence value. \\
				create a child node with the identified dominant behavior. \\
	 			//check with it's parent node \\	
 	 			\If{both nodes satisfy the confidence threshold}
 		    	{ 
 			 	 	
 			 	 	\If{both nodes represent same behavior class}
 	 	 		    	{ 
 	 	 			 	 	mark the child node as `REDUNDANT' node.
 	 	 			 	}
 			 	}
 	
 	 			add a subtree with new node and associated context values.\\
 	 			//recursively do this with remaining contexts \\	
 	 			$AGT(DS_{sub}, \{context \textunderscore list - C_{split} \}, BHs))$	
 	 		}		
 	 	} 
 	 	
 	 	return $N$ \\ 	
 	 \end{algorithm}

\begin{figure}
   	\centering
   	\includegraphics[width=.8\linewidth, height = 5cm]{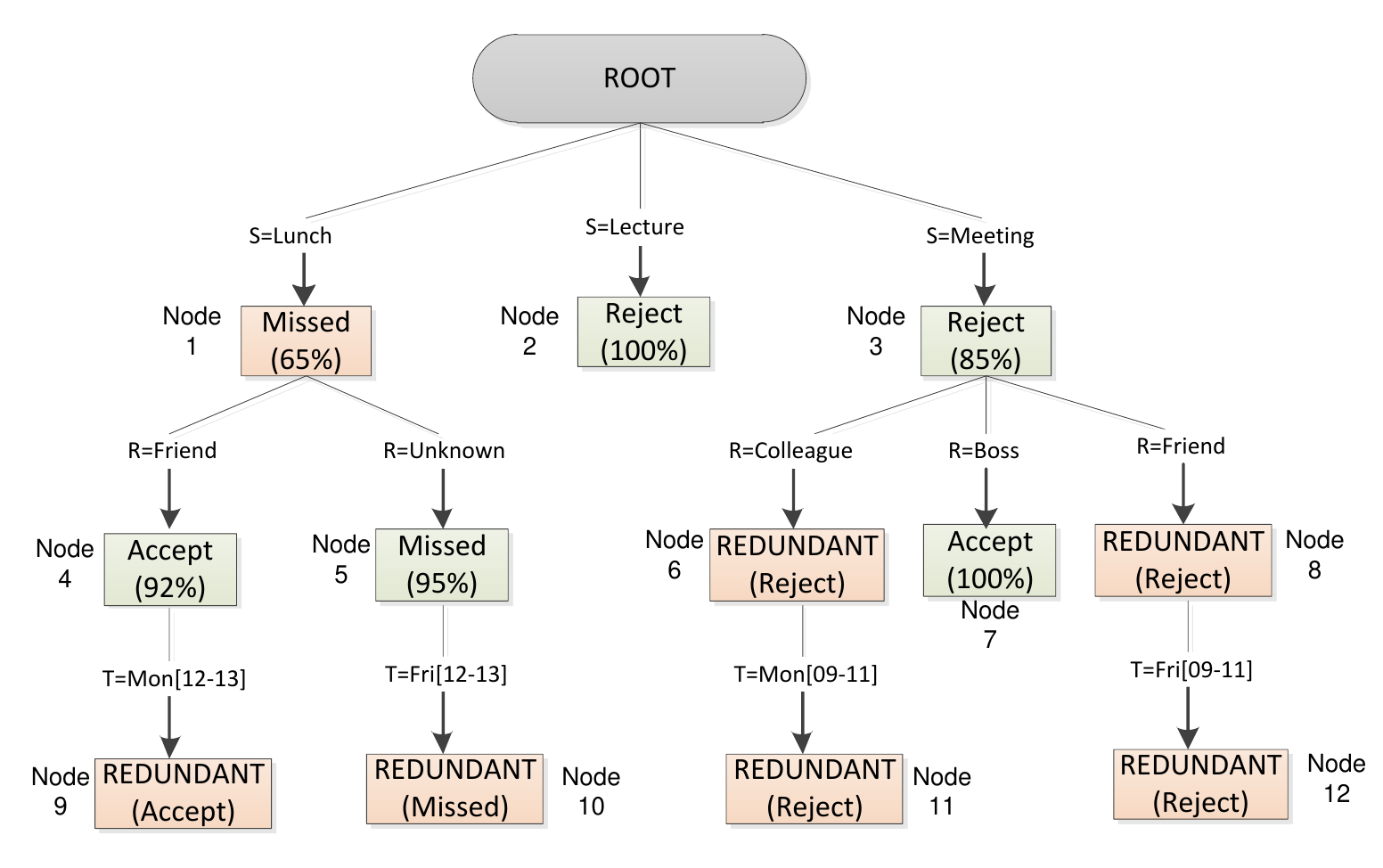}
   	\caption{An example of the tree (AGT) identifying `REDUNDANT' nodes}
   	\label{fig:tree}
   \end{figure}

Figure \ref{fig:tree} shows an example of such an association generation tree containing `REDUNDANT' nodes for the contexts (mentioned above) in phone call behaviors of a user, when the minimum confidence preference is 80\%.
 
 Once the tree has been generated, rules are extracted by traversing the tree. To do this, we first identify the valid rule generating nodes from the tree. \textit{A node is taken into account as a valid rule generating node if it satisfies individual's preferred confidence threshold and not classified as `REDUNDANT' node.} The followings are examples of produced behavioral association rules from the tree.
 
 \noindent $R_1: {Lecture \Rightarrow Reject}$ (conf = 100\%, using Node 2) \\
 $R_2: {Meeting \Rightarrow Reject}$ (conf = 85\%, using Node 3) \\
 $R_3: {Lunch, Friend \Rightarrow Accept}$ (conf = 92\%, using Node 4) \\
 $R_4: {Lunch, Unknown \Rightarrow Missed}$ (conf = 95\%, using Node 5) \\
 $R_5: {Meeting, Boss \Rightarrow Accept}$ (conf = 100\%, using Node 7) \\
  	 
 Rule $R_1$ states that the user always rejects the incoming calls (100\%) when she is in a lecture, which is produced from node 2 in the tree. Similarly, the other non-redundant rules $R_2, R_3, R_4, R_5$ are produced from node 3,4,5, and 7 respectively according to the tree shown in Figure \ref{fig:tree}.\\
 	
 	\section{Experiments}
 	\label{Experiments}
 	In this section, we have conducted experiments on four individual mobile phone users' datasets that consist the phone call records in different contexts. We have implemented our approach in Java programming language and compare the output with the most popular association rule learning technique Apriori \cite{agrawal1994fast}.
 	
 	\subsection{Dataset}
 	We randomly select four individual mobile phone users' datasets from Massachusetts Institute of Technology (MIT) Reality Mining dataset~\cite{eagle2006infering}). These datasets contain three types of phone call behavior, e.g., incoming, missed and outgoing. As can be seen in the dataset, the user's behavior in accepting and rejecting calls are not directly distinguishable in incoming calls in the dataset. As such, we derive accept and reject calls by using the call duration. If the call duration is greater than 0 then the call has been accepted; if it is equal to 0 then the call has been rejected \cite{sarker2016behavior}. The contextual information includes temporal, locational, and social. We also pre-process the temporal data in mobile phone log as it represents continuous time-series with numeric timestamps values (YYYY:MM:DD hh:mm:ss). For this, we use BOTS technique \cite{sarker2017individualized} for producing behavior-oriented time segments, such as Friday[09:00-11:00], Monday[12:00-13:00] etc. Table \ref{Datasets descriptions} describes each dataset of the individual mobile phone user.
 	
 	\begin{table}[htbp!]
 		\centering
 		\caption{Datasets descriptions}
 		\label{Datasets descriptions}
 		\begin{tabular}{|c|c|c|c|} 
 			\hline
 			\bf Datasets & \bf Data Collection Period & \bf Instances \\  
 			\hline
 			Dataset-01 & 5 months & 5119 \\ 
 			\hline
 			Dataset-02 & 3 months & 1229 \\ 
 			\hline
 			Dataset-03 & 4 months & 3255 \\ 
 			\hline
 			Dataset-04 & 4 months & 2096  \\ 
 			\hline
 		\end{tabular}
 	\end{table}

 	\subsection{Evaluation Results}
 	\subsubsection{Effect of Confidence:}
\begin{figure}[htbp!]
 	 	\centering
 	 	\begin{minipage}{.5\textwidth}
 	 	  \centering
 	 	  \includegraphics[width=\linewidth, height = 4cm]{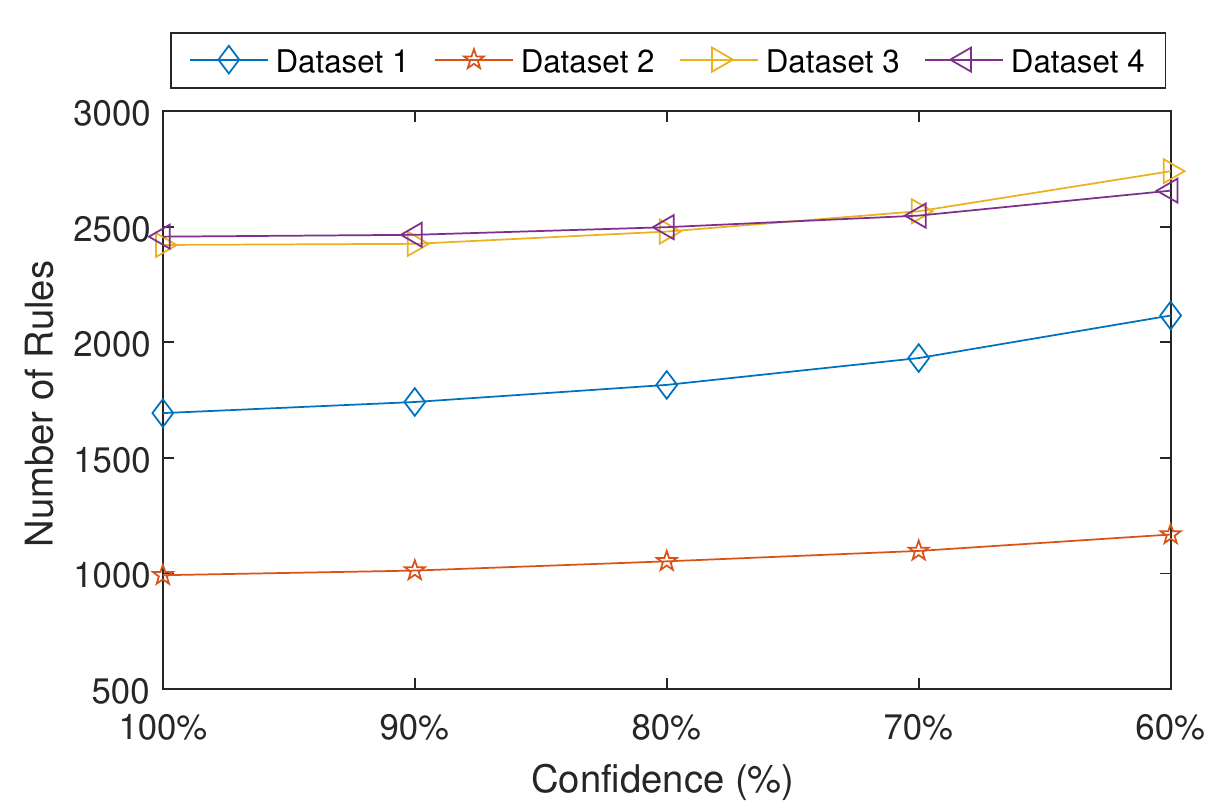}
 	 	  \captionof{figure}{Effect of confidence in ``Apriori"}
 	 	  \label{fig:confidence-effect-apriori}
 	 	\end{minipage}%
 	 	\begin{minipage}{.5\textwidth}
 	 	  \centering
 	 	  \includegraphics[width= \linewidth, height = 4cm]{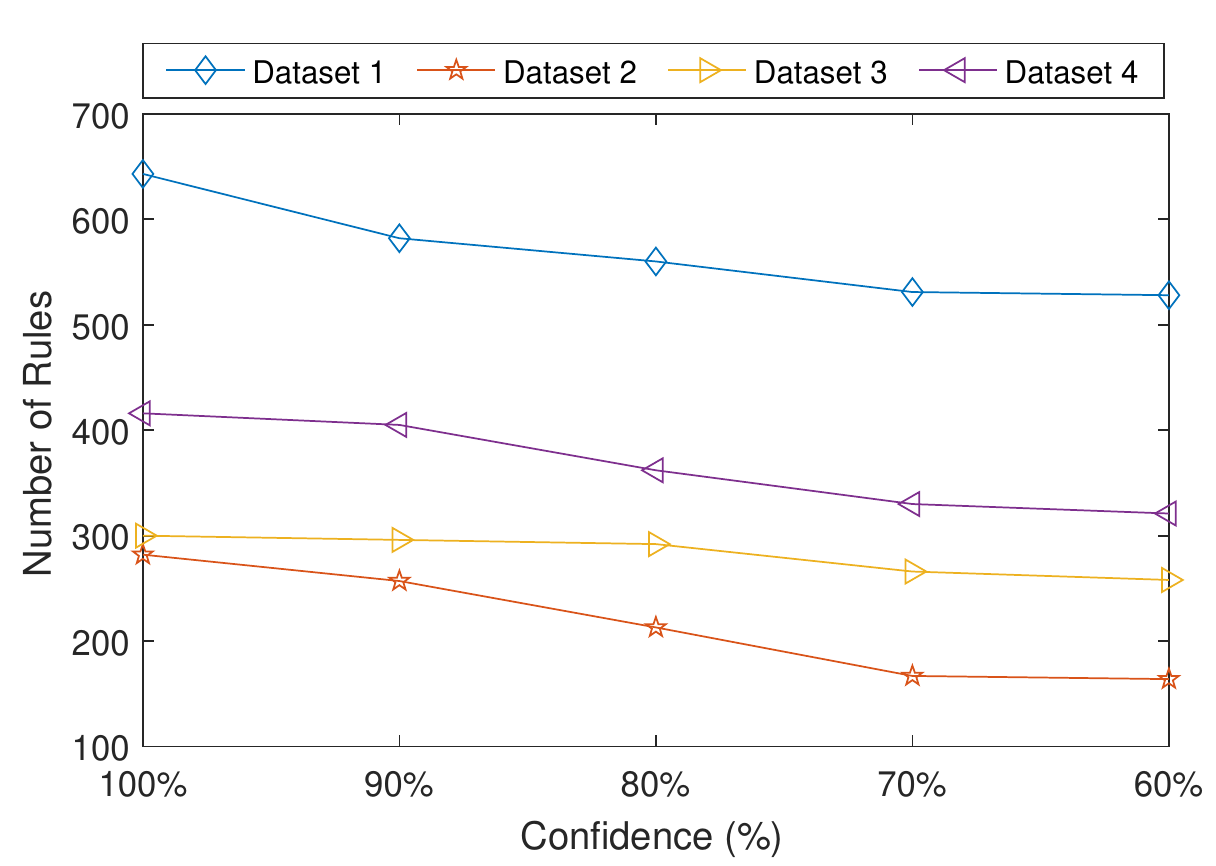}
 	 	  \captionof{figure}{Effect of confidence in our approach}
 	 	  \label{fig:confidence-effect-behavminer}
 	 	\end{minipage}
 	 	\end{figure}

 	In this experiment, we show the effect of confidence on producing behavioral association rules using both approaches. For this, we first illustrate the detailed outcomes by varying the conference threshold from 100\% (maximum) below to 60\% (lowest) for different datasets. Since confidence is directly associated to the \textit{accuracy of rules}, we are not interested to take into account below 60\% as confidence threshold. To show the effect of confidence, Figure \ref{fig:confidence-effect-apriori} and Figure \ref{fig:confidence-effect-behavminer} show the comparison of rule production for different confidence thresholds (accuracy level) for different datasets.
 	
 	If we observe Figure \ref{fig:confidence-effect-apriori}, we see that the produced number of association rules using existing Apriori algorithm \cite{agrawal1994fast} increases with the decrease of confidence threshold. The reason is that it simply takes into account all combination of contexts while producing rules. Thus, for a lower confidence value, it satisfies more associations, and as a result, the output becomes larger. On the other hand, the produced number of behavioral association rules using our technique decreases with the decrease of confidence threshold, shown in Figure \ref{fig:confidence-effect-behavminer}. The main reason is that - for lower confidence threshold, more number of child nodes subsume in their parent node because of creating generalized nodes with the dominant behavior, and as a result, the number of produced rules decreases.
 	
 	\subsubsection{Effectiveness Analysis:}
\begin{figure}[htbp!]
 	\centering
 	\begin{minipage}{.5\textwidth}
 	  \centering
 	  \includegraphics[width=\linewidth, height = 4cm]{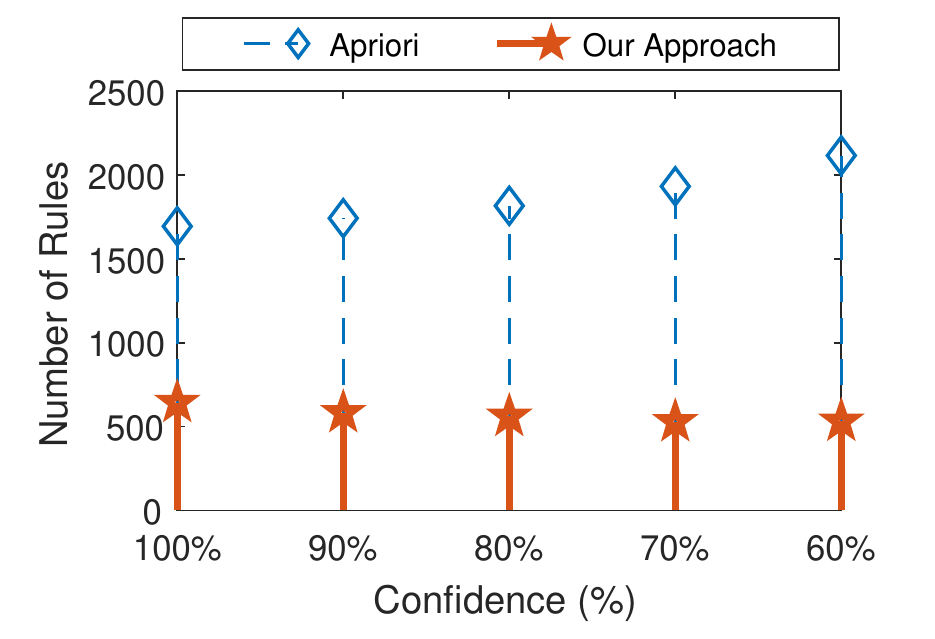}
 	  \captionof{figure}{Utilizing dataset 01}
 	  \label{fig:RM-04}
 	\end{minipage}%
 	\begin{minipage}{.5\textwidth}
 	  \centering
 	  \includegraphics[width= \linewidth, height = 4cm]{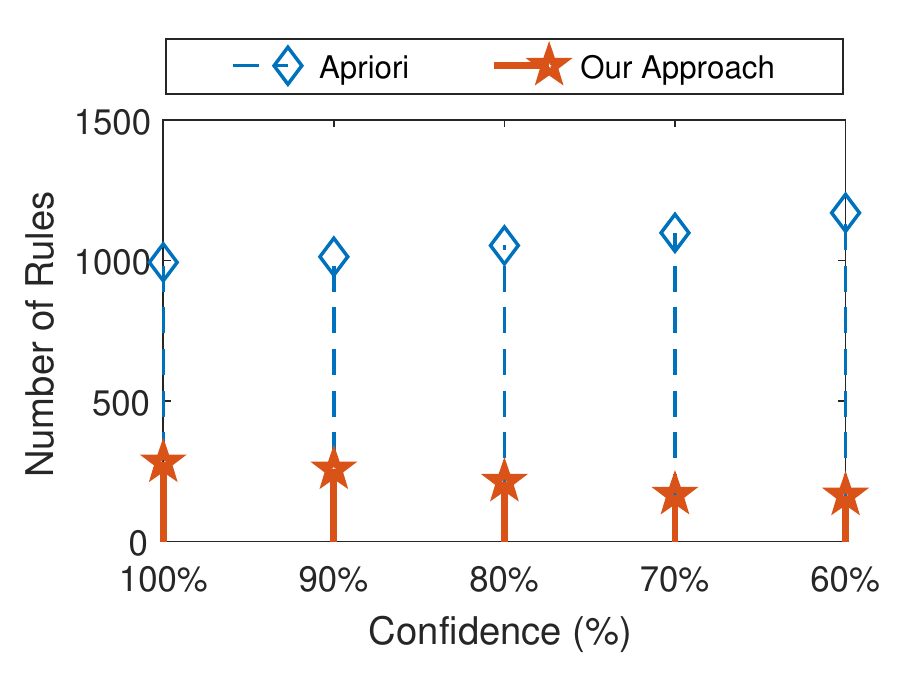}
 	  \captionof{figure}{Utilizing dataset 02}
 	  \label{fig:RM-23}
 	\end{minipage}
 	\end{figure}

\begin{figure}[htbp!]
 	 	\centering
 	 	\begin{minipage}{.5\textwidth}
 	 	  \centering
 	 	  \includegraphics[width=\linewidth, height = 4cm]{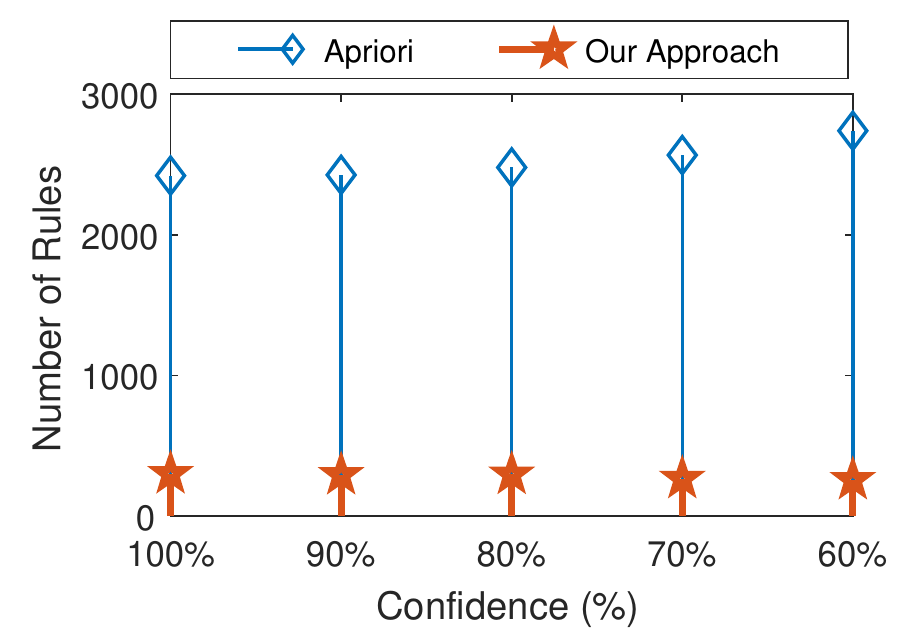}
 	 	  \captionof{figure}{Utilizing dataset 03}
 	 	  \label{fig:RM-26}
 	 	\end{minipage}%
 	 	\begin{minipage}{.5\textwidth}
 	 	  \centering
 	 	  \includegraphics[width= \linewidth, height = 4cm]{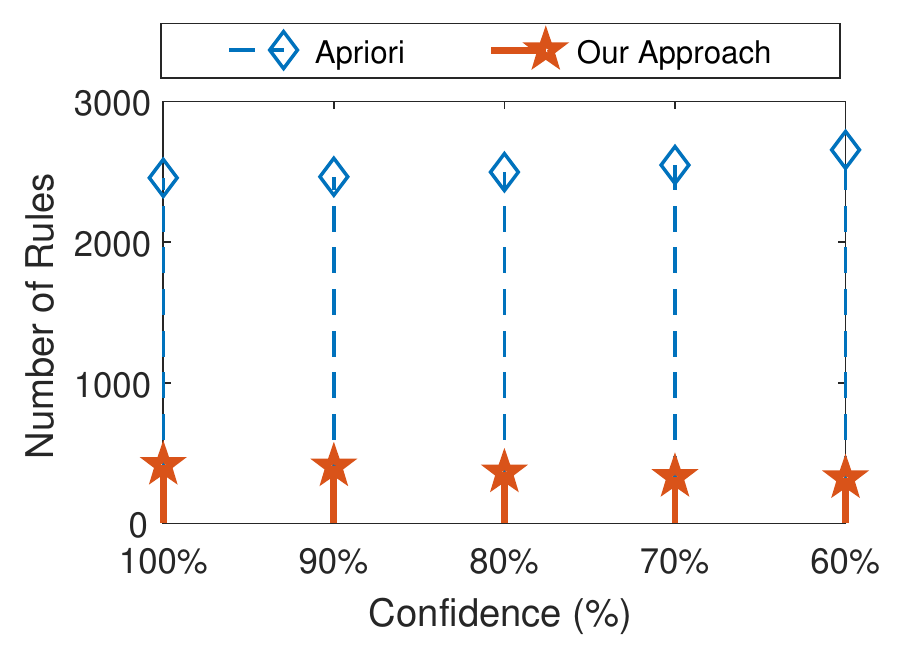}
 	 	  \captionof{figure}{Utilizing dataset 04}
 	 	  \label{fig:RM-51}
 	 	\end{minipage}
 	 \end{figure}

 	To show the effectiveness of our approach, Figure \ref{fig:RM-04}, Figure \ref{fig:RM-23}, Figure \ref{fig:RM-26} and Figure \ref{fig:RM-51} show the relative comparison of produced number of rules for dataset-01, dataset-02, dataset-03 and dataset-04 respectively. For each approach, we use minimum support 1 (one instance) because no rules can be produced below this support \cite{sarker2016behavior}. Moreover, we have explored different confidence threshold, i.e., 100\% (maximum) below to 60\%.
 
 From Figure \ref{fig:RM-04}, Figure \ref{fig:RM-23}, Figure \ref{fig:RM-26} and Figure \ref{fig:RM-51}, we find that our approach significantly reduces the number of extracted rules comparing with traditional association rule learning algorithm for different confidence thresholds. The main reason is that existing approach Apriori \cite{agrawal1994fast} does not take into account redundancy analysis while producing rules and makes the rule-set unnecessarily larger. On the other-hand, we identify and eliminate the redundancy while producing rules and discovers only the non-redundant behavioral association rules. As a result, it significantly reduces the number of rules for a particular confidence threshold for each dataset.
 
\section{Conclusion and Future Work}
\label{Conclusion and Future Work}
 	In this paper, we have presented an approach to effectively identify the redundancy in association rules and to extract a concise set of behavioral association rules which are non-redundant, in order to model phone call behavior of individual mobile phone users. Although we choose phone call contexts as examples, our approach is also applicable to other application domains. We believe that our approach opens a promising path for future research on extracting behavioral association rules of mobile phone users. 
 	
 	In future work, we plan to conduct a range of experiments using additional mobile phone  datasets and to use the discovered non-redundant rules in various predictive services. We have also a plan regarding efficiency analysis of our approach to use in real-time applications, in order to provide the personalized services for the end mobile phone users.

\bibliographystyle{plain}
\bibliography{bibfile/association-rule-learners}

\end{document}